\documentclass[final,5p,times,twocolumn]{elsarticle}
\usepackage{amsmath,amssymb,booktabs,microtype,graphicx}
\usepackage[colorlinks=true,citecolor=blue,linkcolor=blue,urlcolor=blue]{hyperref}

\graphicspath{{./figures/}}
\journal{Physics Letters B}
\newcommand{\na}{n\text{--}\alpha}
\newcommand{\dd}{\mathrm{d}}

\hypersetup{pdfauthor={Myeong-Hwan Mun, Jubin Park, Myung-Ki Cheoun, Eunja Ha}}
\begin{document}
\begin{frontmatter}
\title{Neutron--alpha scattering on a quantum processor:
Spin--orbit-split resonances and weak nonresonant channels}
\author[knu]{Myeong-Hwan Mun}
\author[ssu]{Jubin Park\corref{cor1}}
\ead{honolov@ssu.ac.kr}
\author[ssu]{Myung-Ki Cheoun\corref{cor2}}
\ead{cheoun@ssu.ac.kr}
\author[hu]{Eunja Ha}

\cortext[cor1]{Corresponding author}
\cortext[cor2]{Corresponding author}

\address[knu]{Department of Physics, Kyungpook National University, Daegu
41566, Republic of Korea}
\address[ssu]{Department of Physics and Origin of Matter and Evolution of
Galaxies Institute, Soongsil University, Seoul 06978, Republic of Korea}
\address[hu]{Department of Physics and Research Institute for Natural
Sciences,
Hanyang University, Seoul 04763, Republic of Korea}

\begin{abstract}
We reconstruct elastic neutron--alpha phase shifts in all five channels through
$l=2$ from trapped spectra measured on IBM's Boston quantum computer.
Using a four-qubit implementation of the factorized geometric-time RODEO
algorithm, we resolve the spin--orbit-split $p_{3/2}$ and $p_{1/2}$
resonances of $^5$He and the small nonresonant $d$-wave phase shifts. Classically constructed eight-dimensional
effective Hamiltonians are encoded in three system qubits
and probed with a recycled ancilla.
We fit 40 trapped energies per channel at five confinement strengths
(200 energies in total), using a reciprocal effective-range function for the
$d$ waves. The fits yield free-space phase shifts, the elastic cross section,
and both $p$-wave poles. The broad $p_{1/2}$ pole is more sensitive to the
fit order.
On the $0.05\leq E_{\rm cm}\leq6$ MeV grid, the reconstructed phase shifts
differ by less than $0.073^\circ$ from an independent radial calculation for
the same calibrated interaction. The maximum relative difference in the
$l\leq2$ cross section is $0.185\%$.
In separate measurements of ten low-lying levels in three time-separated
blocks, reversing the evolution times and query phase changes the fitted
energies but leaves the ideal ancilla-history probabilities unchanged.
Averaging forward and reverse estimates reduces the pooled root-mean-square
energy error from $49.59$ eV for forward evolution alone to $20.91$ and
$16.37$ eV in two retrospective equal-shot comparisons. The effect varies
by channel.
Histogram tests reject the working noise model for most hardware scans,
so the conditional fit errors do not quantify the total hardware uncertainty.
\end{abstract}

\end{frontmatter}

\section{Introduction}

A quantum calculation of nuclear scattering must connect a discrete energy
spectrum to the continuum scattering amplitude. For a resonant partial wave,
the energy dependence of the amplitude determines the resonance energy and width.
For a weak nonresonant wave, the interaction produces only small shifts of
the trapped levels from their noninteracting values.

Harmonic confinement provides a route from discrete spectra
to free-space scattering. Busch et al.~\cite{Busch1998}
related the energy levels of two particles with a zero-range
interaction in a harmonic trap to their scattering length.
Luu et al.~\cite{Luu2010} extended the use of trapped spectra
to nucleon--nucleon phase shifts with finite-range
interactions.
At finite trap strength, the confining
potential also affects the relative motion within the range
of the interaction.
Zhang~\cite{Zhang2020MERE} incorporated this effect into
a modified effective-range expansion (MERE),
which was subsequently used to extract neutron--alpha and neutron--$^{24}$O
phase shifts from trapped spectra calculated ab initio~\cite{Zhang2020}.
A related oscillator-based approach, the single-state
harmonic oscillator representation of scattering equations
(SS-HORSE), connects eigenenergies in a truncated oscillator
basis to continuum observables, including neutron--alpha
phase shifts and resonance parameters~\cite{Shirokov2016}.

The trapped spectral problem is accessible to quantum
algorithms without explicitly preparing asymptotic
scattering states.
The RODEO algorithm of Choi et al.~\cite{Choi2021} uses
controlled time evolution and ancilla measurements
to filter spectral components near a query energy.
Wang et al.~\cite{Wang2024} combined RODEO, harmonic confinement, and MERE
to extract $s$-wave phase shifts for model neutron--proton
and neutron--alpha scattering from trapped energies obtained
with the IBM Qiskit simulator.
Sharma et al.~\cite{Sharma2024} obtained neutron--proton
${}^{3}S_1$ phase shifts on superconducting hardware
using a variational quantum eigensolver and quantum
subspace expansion. Complementary approaches use real-time
evolution~\cite{Turro2024}, integrated correlation
functions~\cite{Guo2026}, or complex scaling, which has
been implemented on quantum hardware to extract
alpha--alpha resonance energies~\cite{Zhang2025Resonances}.

Our recent neutron--proton studies developed the hardware
implementation of the trapped-spectrum RODEO approach.
Ancilla recycling reduced
the number of active qubits
in a superconducting implementation that preserved the same
ideal ten-cycle filter as a static trapped-ion circuit
used for benchmarking~\cite{MunRecycling2026}.
Factorized geometric-time circuits were then used to
reconstruct free-space $s$-wave phase shifts from
measurements at several levels and confinement
strengths~\cite{MunFG2026}.
Those results showed that a smaller
root-mean-square trapped-energy error
need not improve the phase shift, which also depends on how the errors are
distributed across the spectrum. These $s$-wave benchmarks
leave open how this RODEO-based hardware strategy performs
when the reconstruction must resolve both spin--orbit-split
resonances and weak higher partial waves.

Elastic neutron--alpha scattering provides these features
within a compact cluster problem. For a spin-zero alpha
particle, the channels through $l=2$ are
$s_{1/2}$, $p_{3/2}$, $p_{1/2}$, $d_{5/2}$, and $d_{3/2}$.
The relatively narrow $p_{3/2}$ and broad $p_{1/2}$
resonances are the low-lying spin--orbit partners of
$^5$He.
Over the energy range considered here, the $s$ wave
is repulsive and the $d$ waves are weak and nonresonant.
The $p$ channels test the recovery of resonance structure
and spin--orbit splitting; the $d$ channels test the
resolution of small interaction-induced displacements
from noninteracting trap levels. Resonance energies and
widths defined by poles of the scattering amplitude
can differ from parameters inferred from real-axis
line shapes, particularly for the broad $p_{1/2}$
resonance~\cite{Csoto1997}.
This system therefore tests the reconstructed amplitude
both on the real energy axis and through its analytic continuation
to complex resonance poles.

Here we extend trapped-spectrum RODEO measurements to the spin--orbit-split
$p$ waves and weak nonresonant $d$ waves. To our knowledge, this is the first
reconstruction of all five elastic neutron--alpha phase shifts through $l=2$
from trapped spectra measured with RODEO on a quantum processor.
On IBM's Boston quantum computer, three system qubits and a recycled ancilla
probe eight-dimensional effective Hamiltonians derived from a fixed, calibrated cluster interaction
based on the Kanada--Kaneko--Nagata--Nomoto (KKNN) potential~\cite{Kanada1979,Odsuren2014}.
The production data comprise 200 trapped-energy estimates
at five confinement strengths. We construct the effective Hamiltonians and compile the circuits classically.
The hardware measurements test spectral inference and continuum reconstruction
for this fixed interaction.

Within each channel, we solve the trap quantization condition and fit the
predicted levels directly to the measured energies at all five confinement
strengths. This avoids evaluating a potentially singular quantization
function at the measured energies.
For the weak $d$ waves, we use a reciprocal effective-range
representation that is regular at noninteracting trap
levels. Together, the five channel fits determine the phase shifts, the elastic cross section,
and the complex poles of both $p$-wave resonances.
For the adopted interaction, the reconstructed central
phase shifts agree with an independent radial solution
to within $0.073^\circ$ on the studied energy grid,
$0.05\leq E_{\rm cm}\leq6$ MeV; the elastic cross section
through $l=2$ differs by at most $0.185\%$ on the same grid.
Varying the order of the effective-range expansion
tests the sensitivity of the extracted pole positions,
particularly for the broad $p_{1/2}$ resonance.

A separate experiment on ten low-lying levels probes
systematic spectral shifts by
jointly reversing the evolution times and the query phase, which leaves the ideal probabilities of ancilla measurement histories unchanged.
Averaging forward and reverse estimates reduces the ensemble
root-mean-square error (RMSE) in energy in both
retrospective equal-shot comparisons, with channel-dependent effects.
These diagnostic measurements are analyzed separately
from the production reconstruction.
Histogram tests show discrepancies between the measured distributions and
the working noise model in both data sets. This limits the interpretation
of conditional fit uncertainties even when the central values agree with
the reference.

Sections~2 and 3 describe the Hamiltonians, circuits, and energy measurements.
Sections~4 and 5 present the continuum reconstruction and the evolution-sign
tests, respectively. Section~6 gives our conclusions.

\section{Hamiltonian and effective space}

We describe neutron--alpha scattering with an inert spin-zero
alpha core. In channel $(l,j)$, the Hamiltonian for the reduced
radial wave function is
\begin{equation}
 H_{lj}(w)=
 -C\frac{\dd^2}{\dd r^2}
 +\frac{C\,l(l+1)}{r^2}
 +V_{lj}(r)+\frac{w^2r^2}{4C},
 \label{eq:H}
\end{equation}
where $w=\hbar\omega$,
$C=\hbar^2/(2\mu)=25.94448084$ MeV fm$^2$, and
$\mu=m_nm_\alpha/(m_n+m_\alpha)$.
The external harmonic potential discretizes the continuum
and is absent from the free-space problem. No Coulomb
interaction is included.

The potential $V_{lj}$ contains the parity-dependent central
and spin--orbit Gaussian terms of the KKNN
parametrization~\cite{Odsuren2014}, with
\begin{equation}
 \langle\boldsymbol l\!\cdot\!\boldsymbol s\rangle_{lj}
 =\tfrac12[j(j+1)-l(l+1)-\tfrac34].
\end{equation}
We retain the five uncoupled channels through $l=2$.
In the $s$ wave, exact orthogonality to the Pauli-forbidden
$0s$ oscillator state with length $b_f=1.565$ fm restricts
Eq.~\eqref{eq:H} to the Pauli-allowed subspace.
This nonlocal constraint is applied only to that channel.

Four dimensionless factors scale the even-parity central
strength, odd-parity central strength, odd-parity spin--orbit
strength, and odd-parity range.
They were calibrated to evaluated cross sections from the Evaluated Nuclear Data File (ENDF/B-VIII.0) library~\cite{Brown2018} before the hardware analysis and are held fixed throughout.
The radial functions, Pauli prescription, fitted parameters,
and calibration procedure are specified in Sec.~S1 of the
Supplemental Material (SM).
An independent radial solution for this calibrated KKNN-based interaction
provides the numerical reference. Deviations from it measure the accuracy
of the reconstruction for the adopted interaction.

We construct the parent Hamiltonians in a radial
harmonic-oscillator basis of 768 functions with
$\hbar\Omega=2$ MeV. This basis spacing is distinct from
the confinement energies
$w=0.20$, $0.25$, $0.30$, $0.35$, and $0.40$ MeV.
Increasing the basis from 512 to 768 functions changes
the retained energies by less than $4.2\times10^{-4}$ eV.
For each channel and confinement, an Okubo--Lee--Suzuki
construction~\cite{Okubo1954,SuzukiLee1980} gives an eight-dimensional effective Hamiltonian (D8) that preserves eight
selected positive-energy eigenvalues of the parent Hamiltonian below 8 MeV. The 25 Hamiltonians
are encoded in three system qubits and supply
200 levels for the production calculation.

The effective dimension determines the spectral sampling,
rather than the accuracy of a selected eigenvalue inherited
from the converged parent space. With exact selected energies,
the continuum reconstruction gives a maximum phase-shift
deviation of $0.00194^\circ$ from the radial reference.
This control tests the combined level selection and
reconstruction without hardware or energy-inference errors.
SM Sec.~S2 details the parent-space calculation and level
selection, and Table~S1 compares the D8 and D16 controls.
The D16 calculation uses a different level selection and
therefore does not constitute a nested basis extrapolation.
The eight-dimensional space provides
the required accuracy for this benchmark with one fewer
system qubit. Both the effective-space construction and
the spectral compilation use classical diagonalization.

\section{Spectral interrogation and inference}

For $H_{\rm eff}=U\,\mathrm{diag}(\epsilon_a)U^\dagger$,
an initial basis state $|b\rangle$ has spectral weights
$\rho_a=|U_{ba}|^2$. At query energy $q$,
a RODEO cycle with evolution parameter $t_r$
returns ancilla zero with conditional
probability $z_{ar}=[1+\cos((\epsilon_a-q)t_r)]/2$.
With ancilla measurement and reset between cycles,
the ideal probability of a history
$\boldsymbol h=(h_1,\ldots,h_R)$ is
\begin{equation}
 P(\boldsymbol h|q)
 =\sum_a\rho_a\prod_{r=1}^{R}
 z_{ar}^{1-h_r}(1-z_{ar})^{h_r}.
 \label{eq:joint}
\end{equation}
The shared spectral label correlates the measured rounds,
so their joint distribution cannot generally be replaced
by a product of marginal probabilities.
All attempted shots enter the analysis through the
$2^6$ possible histories, without postselection.

We use six geometrically spaced evolution times $t_r=s\,t_0\eta^{-(r-1)}$,
where $t_0=36.50539374$ MeV$^{-1}$,
$\eta=1.199230706$, and $s$ is channel dependent.
The system is prepared as $U^\dagger|b\rangle$ in the
spectral representation, and the controlled phases are
implemented with uniformly controlled ancilla rotations.
A diagonal system-only phase is omitted because it
commutes with subsequent spectral operations and cancels
from every history probability.
This reduction preserves Eq.~\eqref{eq:joint}, but need
not preserve the final unmeasured system state
(SM Sec.~S3).
Branch enumeration of the
locally compiled circuits at three queries for each
of the 200 templates gives a maximum probability
discrepancy of $1.7\times10^{-12}$.

Production scans on the IBM Boston quantum processing unit (QPU) used nine query energies
per level, with two replicates of 512 shots at each query:
9216 shots per level and 1\,843\,200 shots in total.
The scan windows were chosen classically. We fit the target energy and
its weight, holding the other energies and their relative weights fixed.
The exact target eigenvalue enters the circuit compilation and serves as
the reference for the fitted energy. It is excluded from the likelihood
inputs (SM Sec.~S4). The measurement thus tests local energy estimation
within a previously identified window.

The observation model includes an asymmetric affine
ancilla response and effective transitions between
spectral populations,
$T_{ab}=(1-\gamma)\delta_{ab}+\gamma/D$, with $D=8$.
These terms provide a constrained description of the
measured distributions, not a complete physical noise
model.
A deviance test with 499 parametric-bootstrap refits
assesses model adequacy, while a separate empirical
bootstrap estimates conditional sampling fluctuations;
the likelihood and bootstrap procedures are specified
in SM Sec.~S4.
Of the 200 production fits, 177 have
bootstrap $p<0.01$, indicating widespread model
inadequacy. We retain the fitted energy centers,
but do not interpret their conditional uncertainties
as total hardware errors. We therefore use unweighted energy residuals in the nominal continuum
reconstruction. Systematic shifts in the fitted energies can still affect
the result.

\section{Phase shifts and resonance poles}

For $k=\sqrt{E/C}$ and $b_w=\sqrt{2C/w}$, the harmonic-trap
quantization function is
\begin{equation}
 \mathcal F_l(E,w)=(-1)^{l+1}
 \left(\frac{2}{b_w}\right)^{2l+1}
 \frac{\Gamma\!\left(\frac{2l+3}{4}-\frac{E}{2w}\right)}
      {\Gamma\!\left(\frac{1-2l}{4}-\frac{E}{2w}\right)}.
 \label{eq:trap}
\end{equation}
We account for finite-range confinement effects through
an expansion in $w^2$. For the $s$ and $p$ waves, the fitted
function approaches
$K_{lj}(E)=k^{2l+1}\cot\delta_{lj}(E)$ as $w\to0$.
For the weak, nonresonant $d$ waves, we instead fit
$G_{lj}=1/K_{lj}$ and evaluate $1/\mathcal F_l$ using
the reversed gamma ratio. This reciprocal representation
is regular at noninteracting oscillator levels, where
$\mathcal F_l$ has poles, but cannot be used across
a pole of $\tan\delta_{lj}$.

For either representation, the nominal expansion is
\begin{equation}
 X_{lj}(E,w)=
 \sum_{i=0}^{3}\sum_{m=0}^{1}c^{(lj)}_{im}E^i w^{2m}.
 \label{eq:expansion}
\end{equation}
We fit in the energy domain to avoid transforming noisy
energies through a potentially singular quantization
function. For each trial set of coefficients, predicted
levels are obtained by solving
$X_{lj}(E,w)=\mathcal F_l(E,w)$, or its reciprocal form,
on the branch associated with each measured level.
We determine the coefficients by minimizing
$\sum_n(E_n^{\rm pred}-\widehat E_n)^2$ in a joint fit
to all 40 energies of a channel. Setting $w=0$ then
determines the free-space amplitude. Reference phase
shifts are not used to select the fit.
Branch tracking, phase conventions, and local sensitivity
to energy perturbations are described in SM Sec.~S5.

\begin{figure*}[t]
 \centering
 \includegraphics[width=\textwidth]{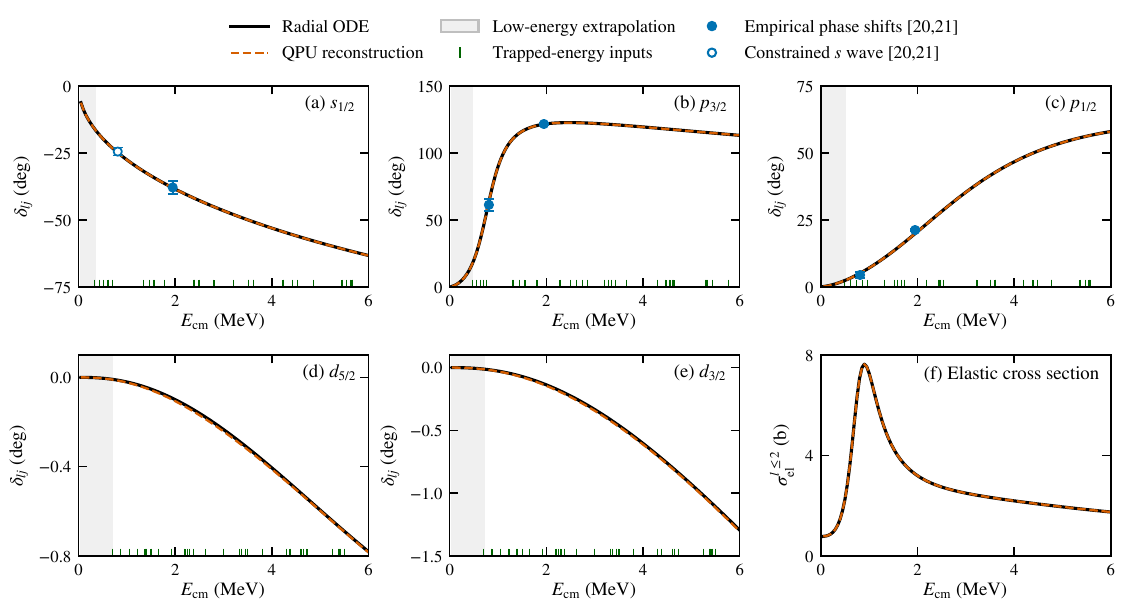}
 \caption{Free-space $\na$ phase shifts and elastic cross
 section reconstructed from the 200-level production data.
 Dashed curves use quantum processing unit (QPU) energy centers; solid curves are
 independent radial solutions of the same calibrated
 interaction. The $s$ and $d$ waves are shown on negative
 phase branches. Grey regions lie below the lowest measured
 trapped energy in each channel; bottom ticks mark trapped
 energies within the displayed interval. Curves are classical
 reconstructions of the measured spectrum.
Symbols in panels (a)--(c) show phase-shift estimates
 from polarized-neutron scattering~\cite{Sawers1968}, with
 error bars taken from the uncertainties reported in
 Ref.~\cite{Sawers1966}. The open symbol in panel (a) denotes
 the constrained lower-energy $s_{1/2}$ value.
 Panel (f) includes
 $l\leq2$ only. No total hardware uncertainty band is assigned.}
 \label{fig:phases}
\end{figure*}

\begin{table}[t]
 \centering
 \caption{Production results by channel.
 $\Delta_{\max}=\max_E|\delta-\delta_{\rm ODE}|$ is evaluated
 on a 120-point grid over $0.05$--$6$ MeV, comparing phases
 modulo $\pi$. ``Exact'' denotes reconstruction from exact
 D8 energies. The last column counts histogram fits with
 bootstrap $p<0.01$. Deviations are benchmark errors,
 not uncertainty estimates.}
 \label{tab:production}
 \small
 \setlength{\tabcolsep}{4pt}
 \begin{tabular}{lrrrr}
 \toprule
 & Energy & \multicolumn{2}{c}{$\Delta_{\max}$ (deg)}
 & Rejected\\
 Channel & RMSE (eV) & Exact & QPU & scans\\
 \midrule
 $s_{1/2}$ & 67.30 & 0.000146 & 0.02511 & 35/40 \\
 $p_{3/2}$ & 70.70 & 0.001933 & 0.07238 & 36/40 \\
 $p_{1/2}$ & 20.72 & 0.000225 & 0.01228 & 36/40 \\
 $d_{5/2}$ & 14.75 & 0.000052 & 0.00871 & 38/40 \\
 $d_{3/2}$ & 14.73 & 0.000041 & 0.01019 & 32/40 \\
 \bottomrule
 \end{tabular}
\end{table}

\begin{figure*}[t]
 \centering
 \includegraphics[width=\textwidth]{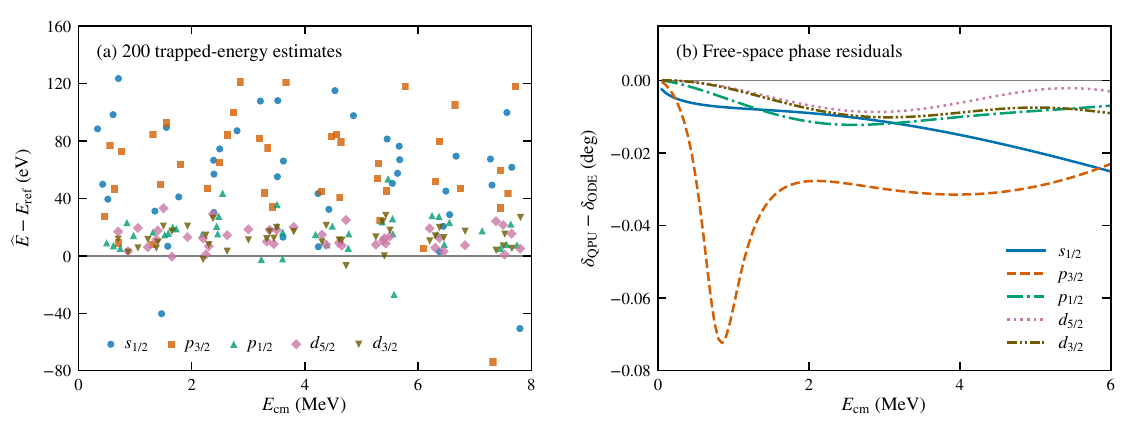}
 \caption{Production residuals for the five channels.
 (a) All 200 fitted QPU energy centers relative to the
 selected parent-space eigenvalues, including fits rejected
 by the goodness-of-fit test.
 (b) Reconstructed QPU phases minus the independent radial
 phases. The different patterns of energy and phase
 residuals reflect the channel-dependent sensitivity
 of the continuum reconstruction.}
 \label{fig:residuals}
\end{figure*}

The reconstructed phases closely follow the radial
reference in Fig.~\ref{fig:phases}; their differences
are resolved in Fig.~\ref{fig:residuals} and summarized
in Table~\ref{tab:production}. The overall production
energy RMSE is $45.59$ eV, with the largest errors
in $s_{1/2}$ and $p_{3/2}$ and values near 15 eV
in the two $d$ waves. The maximum phase deviation is
$0.07238^\circ$ in $p_{3/2}$, while the remaining
channels stay below $0.026^\circ$. Recovery of the
small, negative $d$-wave phases tests the reconstruction
in a regime where the interaction produces only weak
shifts of the trapped levels.

For comparison with experiment, Fig.~\ref{fig:phases}(a)--(c)
 shows the $s_{1/2}$, $p_{3/2}$, and $p_{1/2}$ phase shifts extracted by
 Sawers et al.~\cite{Sawers1968} at incident neutron energies of
 $1.015$ and $2.440$~MeV, corresponding to $E_{\rm cm}=0.811$ and
 $1.949$~MeV, respectively. The error bars use the uncertainties
 reported in Sawers's dissertation~\cite{Sawers1966}. The $s$-wave
 values are shifted by $-180^\circ$ to match the phase branch used
 for the curves. The lower-energy $s_{1/2}$ value was fixed using
 the hard-sphere radius inferred at the higher energy~\cite{Sawers1968}.

For an unpolarized neutron and a spin-zero target,
the elastic cross section through $l=2$ is
\begin{equation}
 \sigma_{\rm el}^{l\leq2}(E)=\frac{4\pi}{k^2}
 \sum_{l=0}^{2}
 \left[(l+1)\sin^2\delta_{l+}
       +l\sin^2\delta_{l-}\right],
 \label{eq:sigma}
\end{equation}
where $\delta_{l\pm}$ denotes the phase shift for $j=l\pm\tfrac12$.
The cross section reconstructed from the QPU energies differs from the corresponding
radial result by at most $0.185\%$ on the displayed grid.
Independent angular integration using the spin amplitudes
defined in SM Sec.~S6 reproduces Eq.~\eqref{eq:sigma}
to numerical precision.
Adding classical $l=3,4$ contributions changes the
cross section by less than $0.0066\%$ over the same
range. This checks the partial-wave truncation for
the adopted interaction, not its physical uncertainty.

We extract the $p$-wave poles by continuing the fitted
effective-range function to complex momentum and solving
$K_{1j}(Ck^2)-ik^3=0$ for
$\operatorname{Re}k>0$ and $\operatorname{Im}k<0$.
The pole energy $E_p=Ck^2=E_R-i\Gamma/2$ defines the resonance energy
$E_R$ relative to the $n+\alpha$ threshold and the width $\Gamma$.
Table~\ref{tab:poles} compares the reconstructed poles
with an independent outgoing-wave radial calculation
described in SM Sec.~S6.
The empirical $R$-function results in the table use a channel radius of \(a=3.3\) fm~\cite{BondFirk1977,Tilley2002}.
The nominal QPU results reproduce both the relatively
narrow $p_{3/2}$ pole and the broad $p_{1/2}$ pole.
These parameters characterize poles of the model
amplitude; they need not coincide with cross-section
maxima or Breit--Wigner parameters, particularly
for the broad resonance~\cite{Csoto1997}.

\begin{table}[t]
  \centering
  \caption{Nominal $p$-wave pole parameters in MeV,
  with literature results for comparison.
  ``Exact energies'' and ``QPU energies'' denote the
  respective reconstructions. All calculations in the
  first three rows use the same fixed interaction.
  Printed digits do not imply hardware confidence intervals.
  Literature results are taken from empirical analyses
  and microscopic calculations. }
  \label{tab:poles}

  \footnotesize
  \setlength{\tabcolsep}{2pt}
  \renewcommand{\arraystretch}{1.12}

\begin{tabular}{
  @{}l@{\hspace{20pt}}cccc@{}
}
    \toprule
    & \multicolumn{2}{c}{$p_{3/2}$}
    & \multicolumn{2}{c}{$p_{1/2}$} \\
    \cmidrule(lr){2-3}
    \cmidrule(l){4-5}
    Method & $E_R$ & $\Gamma$ & $E_R$ & $\Gamma$ \\
    \midrule
    Radial ODE
    & 0.76911 & 0.65114 & 2.00828 & 5.42406 \\
    Exact energies
    & 0.76910 & 0.65111 & 2.00852 & 5.42431 \\
    QPU energies
    & 0.76948 & 0.65145 & 2.00941 & 5.42371 \\
    \midrule
    Empirical $R$-function \cite{BondFirk1977,Tilley2002}
    & 0.77 & 0.64 & 1.97 & 5.22 \\

    Extended $R$-matrix \cite{Csoto1997}
    & 0.80 & 0.65 & 2.07 & 5.57 \\

    Microscopic RGM \cite{Csoto1997}
    & 0.76 & 0.63 & 1.89 & 5.20 \\

    SS-HORSE, JISP16 \cite{Shirokov2018}
    & 0.89 & 0.99 & 1.86 & 5.46 \\

    SS-HORSE, Daejeon16 \cite{Shirokov2018}
    & 0.68 & 0.52 & 2.45 & 5.07 \\
    \bottomrule
  \end{tabular}
\end{table}

To test the continuation, we vary the maximum power of
$E$ from two to four and that of $w^2$ from one to two,
retaining all resulting fits. The QPU $p_{3/2}$ poles
span $E_R=0.76941$--$0.77026$ MeV and
$\Gamma=0.65106$--$0.65201$ MeV.
For $p_{1/2}$, the ranges are
$E_R=2.00731$--$2.02015$ MeV and
$\Gamma=5.37016$--$5.42822$ MeV.
Phase-reconstruction comparisons are given in SM Sec.~S5,
Table~S2, and Figs.~S2 and S3; the pole-order dependence
is shown in SM Sec.~S6 and Fig.~S4.
These ranges measure the sensitivity to the fit order. The width of the
broad $p_{1/2}$ resonance is more sensitive to the continuation, even when
the phase shifts agree closely with the radial reference on the real axis.

\section{Evolution-sign test and repeated measurements}

We test the stability of the inferred energies under
$t_r\to-t_r$, reversing the query phase as well.
This transformation leaves Eq.~\eqref{eq:joint} unchanged,
so forward and reverse circuits have identical ideal
history distributions but may respond differently to
hardware errors. A third arm uses $t_r/2$ with the same
number of cycles. The ten targets are the lowest selected
level in each channel at $w=0.20$ and $0.40$ MeV.
We reuse the forward native templates and compile the
other arms with the same initial layout.
All 30 templates use four active physical qubits and
56 controlled-$Z$ gates, with depths of 272--294.
Halving the evolution times therefore does not halve
the native circuit cost.

Three blocks, r00--r02, were executed on the same day,
with successive execution windows separated by
approximately 174 and 63 minutes. Each arm receives
9216 shots per target and block, giving 829\,440 shots
in 18 jobs. The stored two-qubit and readout calibration
snapshots are unchanged across the blocks; these are
time-separated repeats, not demonstrated independent
calibration epochs. All channels also share a physical
ancilla, so the 30 target--block pairs cannot be treated
as independent device realizations.
Circuit settings are specified in SM Sec.~S3, and execution
timing and calibration records are detailed in SM Sec.~S7.

We define the sign average and contrast as
\begin{equation}
 \widehat E_{\rm av}
 =\frac{\widehat E_++\widehat E_-}{2},
 \qquad
 \widehat E_{\rm odd}
 =\frac{\widehat E_+-\widehat E_-}{2}.
 \label{eq:mean}
\end{equation}
Here $\widehat E_+$ and $\widehat E_-$ denote the forward and reverse energy estimates, respectively.
The average cancels a strictly sign-odd bias without
using the reference energy, but offers no such protection
against a sign-even contribution. The contrast measures
the response to reversal without identifying a unique
microscopic error mechanism.

\begin{figure*}[t]
 \centering
 \includegraphics[width=\textwidth]{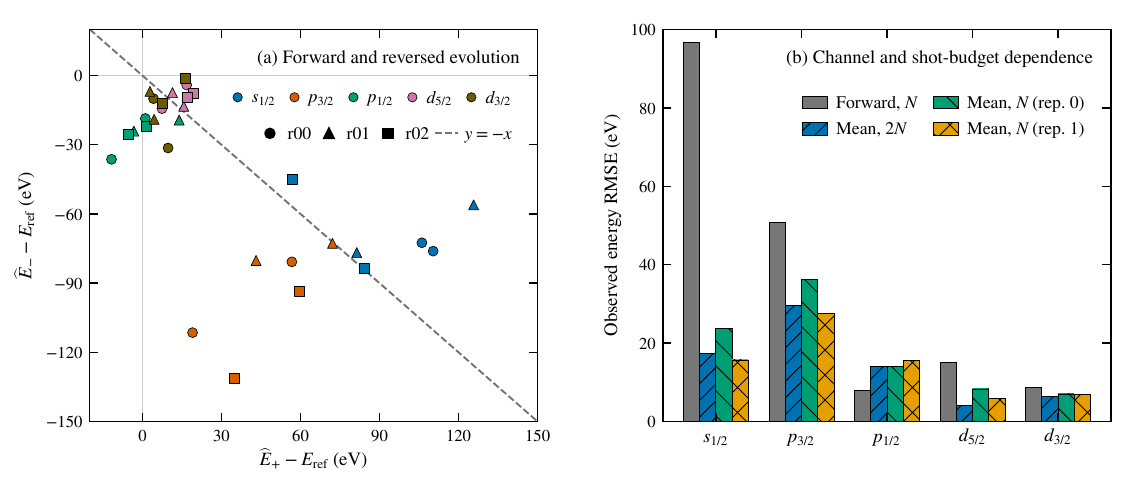}
 \caption{Evolution-sign tests for ten targets in three
 blocks. (a) Forward and reverse energy residuals.
 Colours identify channels as in Fig.~\ref{fig:residuals};
 circles, triangles, and squares denote r00, r01, and r02.
 The dashed line indicates exact cancellation of the
 residuals. (b) Channel RMSEs for forward evolution,
 the full sign average, and both equal-shot partitions.
 With $N=9216$ shots per target and block, the forward
 estimate uses $N$ shots and the full average uses $2N$.
 Each equal-shot partition combines a 4608-shot forward
 fit with the corresponding 4608-shot reverse fit.
 The partitions share the parent data and are not
 independent experiments.}
 \label{fig:sign}
\end{figure*}

All 30 reverse estimates lie below their forward
counterparts [Fig.~\ref{fig:sign}(a)].
The pooled forward and reverse biases are $+32.62$
and $-42.08$ eV, giving an average bias of $-4.73$ eV.
The RMSE decreases from $49.59$ eV for forward evolution
to $16.93$ eV for the full sign average, which uses
twice as many shots. For an equal-shot comparison,
we separately average the forward and reverse fits
from replicate 0 and from replicate 1.
At 9216 shots per target and block, replicates 0 and 1 give RMSEs of
$20.91$ and $16.37$ eV, respectively. We report both retrospective
partitions of the same data.
Channelwise equal-shot results are listed in SM Table~S4.
The improvement is channel dependent [Fig.~\ref{fig:sign}(b)].
The full average reduces
the $s_{1/2}$ RMSE from $96.74$ to $17.46$ eV,
but increases the $p_{1/2}$ RMSE from $7.91$ to
$14.09$ eV. Individual absolute errors improve
in 21 of 30 cases.
The averaged estimates have block RMSEs of $19.20$,
$13.48$, and $17.60$ eV (SM Table~S3 and Fig.~S5(a));
pooling the two later blocks gives $15.67$ eV.
The half-time arm has an RMSE of $133.22$ eV
(SM Table~S5).
Thus, shorter evolution is not beneficial in this
implementation, because it weakens spectral discrimination at the same entangling-gate count.
The sign average provides useful cancellation for
the measured ensemble, but not a uniform correction
across channels.

The histogram test rejects 72 of 90 hardware fits,
compared with 2, 3, and 2 for the ideal, gate-noise,
and backend-calibrated simulators, respectively
(SM Table~S5).
The corresponding energy RMSEs are compared in
SM Fig.~S5(b).
Optimizer convergence does not resolve this model
inadequacy: fitted centers can remain close to the
reference while the predicted distributions disagree
with the observations.
A follow-up scan illustrating the history residuals
is shown in SM Fig.~S1.
The reported RMSEs are therefore
benchmark errors, not hardware standard deviations
or confidence intervals. Moreover, these measurements
sample only the lowest selected levels at two trap
strengths. Establishing an improvement in phase shifts
or resonance poles requires extending the test to
the complete spectral design.

\section{Conclusions}

Our four-qubit calculation on IBM's Boston quantum computer extends the
trapped-spectrum RODEO approach from $s$-wave benchmarks
to all five channels through $l=2$, including the
spin--orbit-split $p$-wave resonances and weak
nonresonant $d$ waves.
The two $p$ channels resolve the spin--orbit partners in $^5$He, while
the weak $d$ waves probe small shifts from the noninteracting trap levels.

Joint energy-domain fits to
40 trapped-energy estimates per channel (200 in total)
at five confinement strengths, together with a reciprocal
effective-range representation for the weak $d$ waves,
connect the same spectral data to phase shifts,
the elastic cross section, and both $p$-wave resonance poles.
For the fixed, calibrated KKNN-based interaction,
the reconstructed central phase shifts and the
$l\leq2$ elastic cross section differ from independent
radial results by at most $0.073^\circ$ and $0.185\%$,
respectively, on the studied grid over
$0.05\leq E_{\rm cm}\leq6$ MeV.
The nominal reconstruction also reproduces both $p$-wave poles. The broad
$p_{1/2}$ pole is more sensitive to the truncation orders in $E$ and $w^2$.

Separate evolution-sign measurements of ten low-lying levels
probe systematic spectral shifts by reversing
both the evolution times and the query phase, which leaves
the ideal ancilla-history probabilities unchanged.
Averaging forward and reverse estimates reduces the
ensemble RMSE in energy in both retrospective equal-shot
comparisons, with channel-dependent effects.
Whether sign averaging improves the phase shifts or pole positions remains
untested for the full 200-level spectrum.
Histogram tests further reveal substantial discrepancies
between the measured distributions and the working
noise model. The agreement of reconstructed central
values therefore does not establish the validity
of the inferred hardware uncertainties.

We construct the effective Hamiltonians classically
and use their eigendecompositions to compile the quantum circuits,
defining a controlled benchmark
for a specified cluster interaction.
A precision determination of the resonance parameters will require repeated
measurements of complete spectra, a statistical model that reproduces the
measured ancilla histories, and control of the analytic continuation,
particularly for the broad $p_{1/2}$ pole.

\section*{Data availability}
The raw QPU counts, job identifiers, transpiled circuits, processed
data, and analysis scripts supporting this work are available
from the corresponding authors upon reasonable request.

\section*{Acknowledgments}
This work was supported by the National Research Foundation of Korea (NRF)
through grants funded by the Korean government (Ministry of Science and ICT,
MSIT): Grant Nos. RS-2026-25487837 and RS-2018-NR031074 (M.-H.M.); Grant
Nos. RS-2021-NR060129, RS-2024-00460031, and RS-2025-16071941 (M.-K.C.); and
Grant No. RS-2025-00513410 (E.-J.H.). Additional support was provided
through NRF Grant Nos. RS-2025-24533596 and RS-2025-25400847, funded by MSIT
and the Ministry of Education, respectively (J.-B.P.).
Quantum computing cloud resources were provided by the Korea Institute of
Science and Technology Information (KISTI) through an NRF grant funded by
the Korean government (MSIT) (Grant No. RS-2025-24534879).
This research was also supported by the “Quantum Information Science R\&D
Ecosystem Creation” program through the NRF, funded by the Korean government
(MSIT) (Grant No. 2020M3H3A1110365).

\end{document}


\begin{frontmatter}
\title{Supplemental Material for
``Neutron–alpha scattering on a quantum processor: Spin–orbit-split resonances and weak nonresonant channels''}
\author[knu]{Myeong-Hwan Mun}
\author[ssu]{Jubin Park\corref{cor1}}
\ead{honolov@ssu.ac.kr}
\author[ssu]{Myung-Ki Cheoun\corref{cor2}}
\ead{cheoun@ssu.ac.kr}
\author[hu]{Eunja Ha}

\cortext[cor1]{Corresponding author}
\cortext[cor2]{Corresponding author}

\address[knu]{Department of Physics, Kyungpook National University, Daegu
41566, Republic of Korea}
\address[ssu]{Department of Physics and Origin of Matter and Evolution of
Galaxies Institute, Soongsil University, Seoul 06978, Republic of Korea}
\address[hu]{Department of Physics and Research Institute for Natural
Sciences,
Hanyang University, Seoul 04763, Republic of Korea}
\end{frontmatter}

This document specifies the interaction, effective-space construction,
ancilla observable, likelihood, and offline reconstruction used in the Letter.
The production measurement covers 200 levels in one hardware block. The
evolution-sign test repeats ten low-lying levels in three time-separated blocks.
All energies are center-of-mass energies unless otherwise stated.

\section{Interaction and Pauli projection}
The calculation is a Hermitian two-cluster model with a spin-zero alpha
particle and a spin-$1/2$ neutron. The masses and conversion factor are
\begin{align}
 m_n c^2&=939.56542194~\mathrm{MeV},\\
 m_\alpha c^2&=3727.3794118~\mathrm{MeV},\\
 \hbar c&=197.3269804593025~\mathrm{MeV\,fm},\\
 C&=25.944480838681724~\mathrm{MeV\,fm^2}.
\end{align}
The alpha mass includes its binding energy.
No electromagnetic or absorptive optical-potential term is added.
Excited-core channels and explicit five-body correlations are outside this model.

To specify the local interaction without convention ambiguities, define
$L_{lj}=[j(j+1)-l(l+1)-3/4]/2$ and the four unscaled functions
\begin{align}
 V_c(r)&=-96.3e^{-0.36r^2}+77.0e^{-0.90r^2},\\
 V_\pi(r)&=(-1)^l\big[34.0e^{-0.20r^2}
                  -85.0e^{-0.53r^2}+51.0e^{-2.50r^2}\big],\\
 V_{ls}(r)&=-16.8L_{lj}e^{-0.52r^2},\\
 V_{ls,\pi}(r)&=L_{lj}[1+0.3(-1)^{l-1}]
 \big[-20.0e^{-0.396r^2}+20.0e^{-2.200r^2}\big].
 \label{eq:potential}
\end{align}
Here $r$ is in fm and each function is in MeV.
These are the coefficients of the KKNN potential tabulated in
Ref.~\cite{Odsuren2014}.
For even $l$, the interaction used in the experiment is
$a_e(V_c+V_\pi)+V_{ls}+V_{ls,\pi}$.
For odd $l$, all four functions are evaluated at $r/a_r$, and the
central and spin--orbit sums are multiplied by $a_o$ and $a_s$, respectively.
The calibrated values are
\begin{align}
 (a_e,a_o)&=(1.1326522347059225,\ 0.9346485514492728),\nonumber\\
 (a_s,a_r)&=(0.8988074686257918,\ 1.0319355712314366).
 \label{eq:scales}
\end{align}
The number of digits in Eq.~\eqref{eq:scales} specifies a reproducible
fixed input; it does not represent an experimental parameter precision.

For $l=0$, the forbidden reduced radial state is
\begin{equation}
 \begin{aligned}
 f(r)&=\frac{2r}{\pi^{1/4}b_f^{3/2}}e^{-r^2/(2b_f^2)},\\
 b_f&=1.565~\mathrm{fm},\qquad
 \int_0^\infty |f(r)|^2\dd r=1.
 \end{aligned}
\end{equation}
The Hamiltonian is restricted to the orthogonal complement of $f$.
This implements the infinite-strength limit of the orthogonalizing
pseudopotential $\lambda|f\rangle\langle f|$; no finite, arbitrarily
large penalty is used in the production eigenspectrum.
Only this one $s$-wave component is excluded.
It is a specified cluster-model Pauli prescription, not full
antisymmetrization of five dynamical nucleons.

The calibration record uses ENDF/B-VIII.0 evaluated elastic and angular
cross sections~\cite{Brown2018} at
$E_{\rm cm}=0.15$, $0.35$, $0.60$, $0.80$, $1.0$, $1.4$, $2$, $3$, $4$, $5$
and $6$ MeV, and angles $15^\circ$, $45^\circ$, $75^\circ$, $105^\circ$,
$135^\circ$ and $165^\circ$.
The archived objective is
\begin{equation}
 \begin{aligned}
 Q=\sum_E\Bigg[&
 \left(\frac{\sigma(E)}{\sigma_{\rm eval}(E)}-1\right)^2\\
 &+\frac{1}{N_\theta}\sum_\theta
 \left(\frac{\sigma_\Omega(E,\theta)-\sigma_{\Omega,\rm eval}(E,\theta)}
 {\max_\theta\sigma_{\Omega,\rm eval}(E,\theta)}\right)^2\Bigg],
 \end{aligned}
\end{equation}
with $Q=3.6447158814610316\times10^{-4}$.
The calibration used a finite-difference radial solver with $h=0.005$ fm
and partial waves through $l=4$.
The stored holdout energy set is $0.25$, $0.45$, $0.70$, $0.90$, $1.2$,
$1.7$, $2.5$, $3.5$, $4.5$ and $5.5$ MeV.
These holdout points are drawn from the same evaluation and are not
independent raw experiments.
A parameter covariance was not estimated; $Q$ is not a statistical
likelihood and correlated evaluated points cannot be treated as independent
measurements. The paper therefore makes no experimental goodness-of-fit
or interaction-uncertainty claim.

\section{Parent basis and effective Hamiltonians}
A convenient normalized reduced oscillator basis, with
$b=\sqrt{2C/(\hbar\Omega)}$, is
\begin{equation}
 \begin{aligned}
 u_{nl}(r)={}&(-1)^n
 \sqrt{\frac{2n!}{b^{3}\Gamma(n+l+3/2)}}\,
 r\left(\frac r b\right)^l\\
 &\times e^{-r^2/(2b^2)}L_n^{l+1/2}(r^2/b^2).
 \end{aligned}
\end{equation}
The phase convention $(-1)^n$ gives positive off-diagonal elements
of the $r^2$ matrix used by the code.
The production values are $n=0,\ldots,767$, $\hbar\Omega=2$ MeV
and 2600-point generalized-Laguerre quadrature.
The comparison calculation uses 512 basis functions and 2200 points.
The maximum recorded difference of positive levels below 8 MeV is
$4.1510\times10^{-4}$ eV.
Since the basis size and quadrature order change together, this is
a joint numerical convergence check, not an isolated quadrature study.

For the $s$ wave, let $F_n=\langle u_{n0}|f\rangle$ and construct
an orthonormal matrix $B$ spanning $F^\perp$.
The allowed-space diagonalization is
$B^\dagger H(w)B\,v_a=E_a v_a$,
with parent eigenvectors $U_a=Bv_a$.
The other channels require no such restriction.

For D8, eight anchors are distributed uniformly from 0.5 to 7.5 MeV.
A minimum-cost one-to-one assignment selects positive parent levels
below 8 MeV nearest these anchors; distinct anchors cannot select
the same state.
Let $U_s$ collect the selected parent eigenvectors.
Pivoted QR of $U_s^T$ selects eight oscillator rows,
and the resulting square projection has singular-value decomposition
$A=L\Sigma R^\dagger$.
Define $Q=LR^\dagger$ and
\begin{equation}
 H_{\rm eff}=Q\,\mathrm{diag}(E_a)Q^\dagger,
 \qquad W=U_sQ^\dagger.
\end{equation}
Then $W^\dagger W=1$ and the effective spectrum agrees
with the selected parent spectrum.
This is the polar form of a Hermitian effective-space
construction~\cite{Okubo1954,SuzukiLee1980}.
The oscillator rows are selected for conditioning;
they are not simply the first eight oscillator functions.
A rank-deficient projected overlap is rejected.
Independent diagonalization of all 25 D8 matrices reproduces the selected
parent-space energies to within $10^{-10}$ MeV.

The D16 control retains the first sixteen levels, rather than the
D8 anchor-based design. The corresponding reconstruction errors
are listed in Table~\ref{tab:dimension}.
Some D16 selected levels can exceed the 8 MeV analysis cutoff;
the same cutoff is applied when forming its reconstruction inputs.
Both effective dimensions are compared with the independent radial
solution, so neither finite-dimensional result defines the reference.

\begin{table}[t]
\centering
\caption{Maximum phase deviation from the radial ODE on the
$0.05$--$6$ MeV grid, using exact trapped energies, in degrees.
D8 and D16 require four and five qubits including the ancilla.
These are classical controls; this paper reports D8 hardware measurements.}
\label{tab:dimension}
\begin{tabular}{lrr}
\toprule
Channel & D8: spread selection & D16: first-level selection\\
\midrule
$s_{1/2}$ & 0.000146 & 0.000167 \\
$p_{3/2}$ & 0.001933 & 0.000986 \\
$p_{1/2}$ & 0.000225 & 0.000174 \\
$d_{5/2}$ & 0.000052 & 0.000063 \\
$d_{3/2}$ & 0.000041 & 0.000046 \\
\bottomrule
\end{tabular}
\end{table}

\section{Ancilla distribution and circuit reduction}
For $D=2^{n_s}$, diagonalize the effective Hamiltonian classically
and prepare $U^\dagger|b\rangle$ on the $n_s$ system qubits.
For each spectral component $a$, the controlled evolution
is the ancilla phase $\mathrm{diag}(1,e^{-i\epsilon_a t})$.
It factors as
\begin{equation}
 \mathrm{diag}(1,e^{-i\epsilon_a t})
 =e^{-i\epsilon_a t/2} R_z(-\epsilon_a t).
\end{equation}
The first factor is a diagonal operator on the system register.
All following rotations, queries and measurement Kraus operators
are diagonal in the same spectral label.
For any complete ancilla history, omitting these factors changes
only a spectral phase in the conditional system state.
Tracing over the unmeasured system removes that phase.
This proves equality of every history probability, not only
of the all-zero postselection probability.
It does not prove equality of the output system states.

The ancilla sequence in each cycle consists of a Hadamard,
a system-controlled $R_z$, the phase $\mathrm{diag}(1,e^{iqt})$,
a second Hadamard and measurement.
The ancilla is reset after each of the first five cycles.
The three system qubits persist across all six cycles.
The ideal all-zero probability is
\begin{equation}
 P_0(q)=\sum_a \rho_a
 \prod_{r=1}^6\cos^2\!\left[\frac{(\epsilon_a-q)t_r}{2}\right].
 \label{eq:pzero}
\end{equation}
Replacing $t_r$ by $-t_r$ in both the spectral and query rotations
preserves the full ideal distribution.
It need not preserve the error channels of the transpiled circuit.

The six geometric times are
$t_r=36.50539374\,s\,(1.199230706)^{-(r-1)}$ MeV$^{-1}$.
For the ten follow-up targets, the forward scales are
$s=20$, $20$, $80$, $100$ and $100$ in the ordered channels
$s_{1/2}$, $p_{3/2}$, $p_{1/2}$, $d_{5/2}$ and $d_{3/2}$.
The query offsets relative to the stored scan center are
\begin{equation}
 \Delta q=\frac{0.04}{s}
 (-1,-0.7,-0.4,-0.2,0,0.2,0.4,0.7,1)~\mathrm{MeV}.
\end{equation}
The reverse arm uses the negative times and the same queries.
The half-time arm halves $t_r$ while retaining the forward
arm's query grid, shot budget, and number of cycles.
It changes the spectral response and therefore does not
provide a comparison at matched spectral resolution.
The scan centers were inherited from historical synthetic coarse fits.
They are not independently discovered on hardware, nor recentered
using the measured hardware residuals.

The follow-up circuits use physical qubits $\{60,61,62,76\}$ on IBM Boston.
Their initial logical layout is $(62,76,61,60)$,
and the measured ancilla is physical qubit 61.
The 30 compiled templates all contain 56 CZ gates;
depths range from 272 to 294.
The time-reversed and half-time templates use the same initial layout,
but individual gate sequences and depths can differ.
Their comparison therefore does not isolate a single coherent
rotation error with all other implementation details held fixed.

\section{Likelihood and goodness of fit}
The reference spectrum is used in the classical compilation
of the circuits. For inference of the local target $a_*$,
its reference eigenvalue is excluded from the likelihood inputs.
The fitted parameters are the target energy $E_*$, its weight $A$,
two asymmetric response parameters $u,v$, and a
population-refresh parameter $\gamma$.
The non-target energies are fixed, and their relative weights
$\bar\rho_a$ are normalized on $a\ne a_*$, so that
\begin{equation}
 \rho_{a_*}=A,\qquad \rho_{a\ne a_*}=(1-A)\bar\rho_a.
\end{equation}
For spectral state $a$ at cycle $r$, the zero emission probability is
\begin{equation}
 \begin{aligned}
 f_{ar}&=u+(1-u-v)\frac{1+\cos[(\epsilon_a-q)t_r]}{2},\\
 T_{ab}&=(1-\gamma)\delta_{ab}+\gamma/D.
 \end{aligned}
\end{equation}
The joint history probability is obtained by propagating this
hidden spectral population between emissions.
This effective Markov model is a working description of
inter-cycle changes; it is not derived from an IBM gate-level
master equation.
The nominal fits use common response parameters $u,v$
for all six cycles.
The target-energy bound is the scan window; the target weight is
bounded by $10^{-6}\leq A\leq1-10^{-6}$,
$10^{-7}\leq u,v\leq0.49$, and $0\leq\gamma\leq0.5$.

For all counts $n_{qh}$, the minimized negative log likelihood,
up to a parameter-independent constant, is
\begin{equation}
 -\log\mathcal L=-\sum_{q,h}n_{qh}\log P_h(q;\vartheta).
\end{equation}
The fit uses every measured history, including those with a nonzero ancilla
outcome in an early cycle. Fits use three initial center positions and
retain the optimum of the same likelihood.
Statistical covariance is computed only for a fully identified
Fisher matrix. It is not formed by truncating singular directions
and is not rescaled to enforce unit reduced chi-square.
All 90 follow-up hardware fits converge numerically and have full
Fisher rank; this does not establish their statistical validity.

The multinomial deviance is
\begin{equation}
  \mathcal{D}
  = 2\sum_{q,h:\,n_{qh}>0}
    n_{qh}\log\frac{n_{qh}}{N_q\widehat{P}_h(q)}.
\end{equation}
For its parametric-bootstrap test, 499 null data sets are generated
at the fitted probabilities and all parameters are refitted.
With the finite-bootstrap correction,
\(p=[1+\#(\mathcal{D}^{*} \geq \mathcal{D})]/500\),
so the minimum reported value is 0.002.
The primary rejection threshold is $p<0.01$.
A separate 499-draw empirical bootstrap resamples each replicate
at its observed multinomial proportions.
Its intervals remain conditional on the estimator and the measured
block; resampling cannot recover an unknown persistent bias.

\begin{figure*}[t]
\centering
\includegraphics[width=\textwidth]{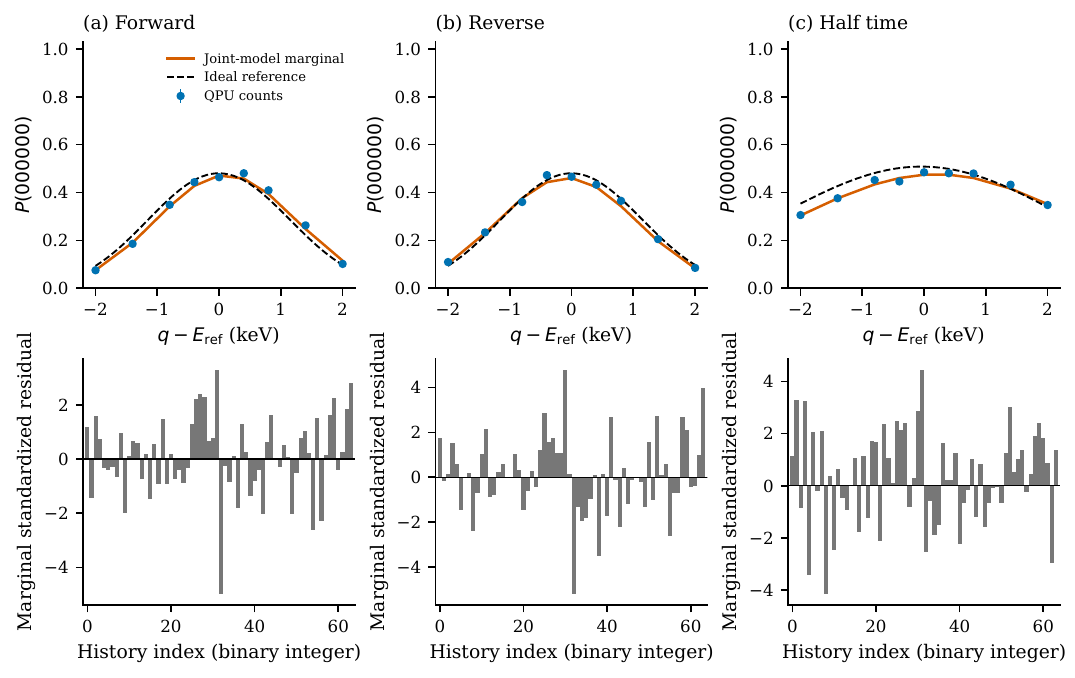}
\caption{Example RODEO scans from the first follow-up block:
$s_{1/2}$, $w=0.20$ MeV, lowest selected level.
The upper panels show the measured all-zero marginal,
its fitted joint-model marginal, and the ideal spectral response
at the reference energy. Error bars are binomial counting errors
of this marginal, not energy uncertainties.
The lower panels compare the counts summed over query points
with the corresponding fitted expectations for all 64 histories.
The residual denominator is
$[\sum_qN_qP_{qh}(1-P_{qh})]^{1/2}$.
Residuals of different histories are correlated.
The three displayed arms are illustrative; all fits enter the reported tests.}
\label{fig:scans}
\end{figure*}

Figure~\ref{fig:scans} illustrates why a satisfactory-looking
postselection peak is insufficient to validate a joint likelihood.
In the follow-up QPU data, the rejection counts are 27/30 for forward,
21/30 for reverse and 24/30 for half-time scans.
The reference lies within the descriptive interval
$\widehat E\pm1.96\sigma_{\rm conditional}$ for only 13/30,
5/30 and 10/30 fits, respectively.
Here \(\sigma_{\mathrm{conditional}}\) is the
empirical-bootstrap standard deviation of the fitted energy.
Shared blocks, shared qubits, and model misspecification limit the
interpretation of these fractions. 

\section{Finite-confinement reconstruction and conditioning}
The Letter's joint ansatz contains eight coefficients per channel.
For $l=0,1$ it describes the finite-confinement extension of
$K=k^{2l+1}\cot\delta$, and for $l=2$ that of $G=1/K$.
On the trap side of the $d$-wave quantization condition, we evaluate the
reciprocal gamma-function ratio directly. It vanishes at the
noninteracting oscillator levels.
All roots are constrained to their observed quantization branches.
A bracketed fallback is used if the local Newton iteration fails.
No measured energy is replaced by an exact eigenvalue and
no target is discarded on the basis of its residual.

The fit coefficients \(c\) are determined by minimizing
\begin{equation}
 S(c)=\sum_{n=1}^{40}
 [E_n^{\rm pred}(c)-\widehat E_n]^2.
\end{equation}
A fixed numerical objective scale affects convergence tolerances only.
It is not assigned as an energy error.
We use unweighted energy residuals because the goodness-of-fit test
rejects the observation model for most hardware scans. We therefore
quote no continuum error band based on the conditional energy covariance.

The reconstructed phase is obtained from
$\delta=\operatorname{atan2}(k^{2l+1},K)$, or
$\delta=\arctan(k^{2l+1}G)$, modulo $\pi$.
Comparisons use
\begin{equation}
 \Delta\delta=\left[(\delta-\delta_{\rm ODE}+\pi/2)\bmod\pi\right]-\pi/2.
\end{equation}
Continuous negative branches for $s,d$ are used in the figures.
This convention changes neither $S_{lj}=e^{2i\delta_{lj}}$
nor any cross section.
It also prevents a phase near $180^\circ$ from being misidentified
as a large positive $d$-wave phase near threshold.

The dependence of a reconstructed phase on the spectrum
is not determined by an energy RMSE.
Let $J_{ni}=\partial E_n^{\rm pred}/\partial c_i$ be evaluated
at the fitted coefficients. If $J$ has full column rank,
the Gauss--Newton approximation gives
\begin{equation}
 \Delta\delta(E)\simeq
 [\nabla_c\delta(E)]^T(J^TJ)^{-1}J^T\Delta\boldsymbol E.
 \label{eq:sensitivity}
\end{equation}
Thus the direction of a correlated energy error, its channel,
and the extrapolation point all matter.
Here $\Delta\boldsymbol E$ denotes a perturbation of the
measured energy inputs. Equation~\eqref{eq:sensitivity}
describes local sensitivity; it does not provide calibrated
uncertainties when the observation model is misspecified.

We evaluate energy orders $d_E=2,3,4$ and trap orders $d_w=1,2$,
where $d_w=2$ includes $w^4$.
All six alternatives are retained, including the primary $(3,1)$ fit.
We also remove each trap in turn, fit the other four,
and predict the omitted levels on their identified branches.
These tests diagnose dependence on the global polynomial assumption.
They do not measure its truncation probability distribution.

\begin{table*}[t]
\centering
\caption{Reconstruction comparisons for the QPU production energies, in degrees.
The Two-stage \(K\) and Joint primary columns give maximum
deviations from the radial ODE. The Order variation and Trap removal columns
give maximum changes from the primary fit.
The two-stage method uses a quadratic \(K\) fit followed by trap extrapolation.
Both methods use \(K\) for \(p_{3/2}\), with different fit orders and procedures.
The primary fit uses \(G=1/K\) for the \(d\) waves.
For \(p_{3/2}\), the primary fit deviates from the radial reference by up to
\(0.07238^\circ\). Varying the fit order or removing one trap changes the
phase by up to \(0.10477^\circ\) and \(0.03987^\circ\), respectively,
relative to the primary fit.
These changes diagnose reconstruction sensitivity and are not uncertainty bounds.
The two-stage fit gives a smaller reference deviation
for \(p_{3/2}\).}
\label{tab:reconstruction}
\begin{tabular}{lrrrr}
\toprule
Channel & Two-stage $K$ & Joint primary & Order variation & Trap removal\\
\midrule
$s_{1/2}$ & 0.03050 & 0.02511 & 0.05259 & 0.01489 \\
$p_{3/2}$ & 0.03880 & 0.07238 & 0.10477 & 0.03987 \\
$p_{1/2}$ & 0.06757 & 0.01228 & 0.05203 & 0.00789 \\
$d_{5/2}$ & 0.02164 & 0.00871 & 0.01511 & 0.00318 \\
$d_{3/2}$ & 0.01717 & 0.01019 & 0.00756 & 0.00248 \\
\bottomrule
\end{tabular}
\end{table*}

\begin{figure*}[t]
\centering
\includegraphics[width=\textwidth]{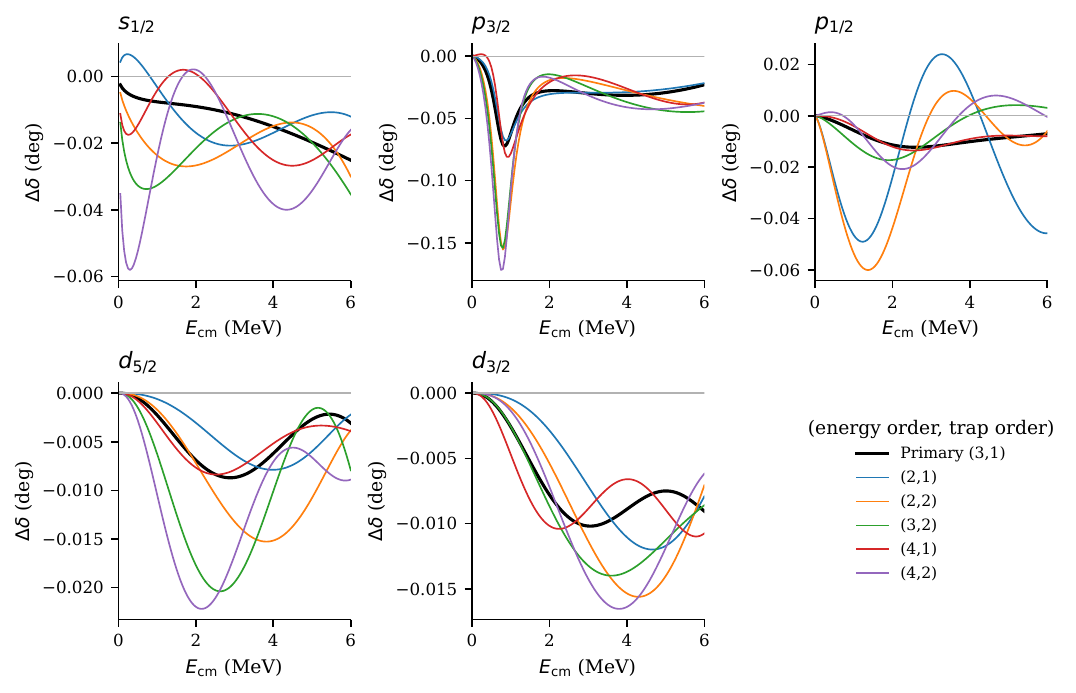}
\caption{Phase residuals relative to the fixed-model radial solution
for all six declared polynomial orders. The primary $(3,1)$ result
is black. The spread is a fit-form diagnostic and is not a confidence band.}
\end{figure*}
\begin{figure*}[t]
\centering
\includegraphics[width=\textwidth]{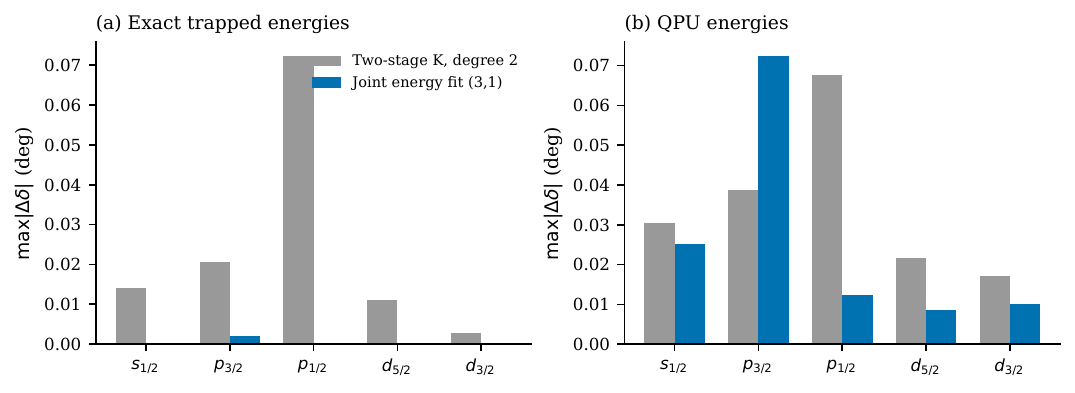}
\caption{Two-stage quadratic $K$ extrapolation and the primary joint
energy-domain reconstruction applied to identical exact or QPU energies.
Both reconstructions use classical postprocessing of the same spectra.}
\end{figure*}

\section{Poles, observables and comparison with nuclear analyses}
The partial-wave amplitude has denominator
$K_{lj}(Ck^2)-ik^{2l+1}$.
We retain polynomial roots with $\operatorname{Re}k>0$,
$\operatorname{Im}k<0$, $0<E_R<6$ MeV and $0<\Gamma<12$ MeV.
For a reciprocal fit the corresponding equation is
$1-ik^{2l+1}G_{lj}(Ck^2)=0$.
Apparent $d$-wave roots of this polynomial outside its
nonresonant domain are not identified as physical resonances.

At real energies, we match radial solutions to free
Riccati--Bessel functions at \(24\,\mathrm{fm}\).
In the \(s\) wave, \(y\) is the regular homogeneous solution and \(z\)
satisfies \((H_{\mathrm{loc}}-E)z=-f\), with \(f\) from Eq.~(S10).
The combination
\(u=y-z\langle f\mid y\rangle/\langle f\mid z\rangle\)
enforces \(\langle f\mid u\rangle=0\).

For the poles, the regular and outgoing solutions
have a zero Wronskian at \(r_{\mathrm{match}}\).
The outer boundary at \(r_{\max}\) uses the free Riccati--Hankel
logarithmic derivative. Changing
\((r_{\max},r_{\mathrm{match}})\) from \((24,5)\) to
\((28,6)\,\mathrm{fm}\) shifts the \(p_{1/2}\) pole by less than
\(1.5\times10^{-11}\,\mathrm{MeV}\) in both parameters,
with smaller shifts for \(p_{3/2}\).

\begin{figure*}[t]
\centering
\includegraphics[width=\textwidth]{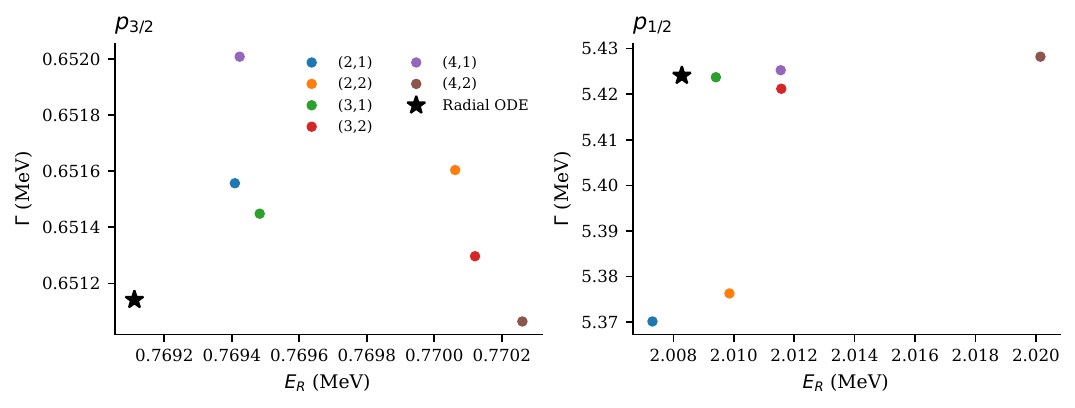}
\caption{\(p\)-wave pole candidates
obtained from the QPU production energies for
all six reconstruction orders. The star is the independent radial pole
of the same interaction. The width of the broad \(p_{1/2}\) state is more
sensitive to the polynomial continuation. No error bars are inferred
from this spread.}
\end{figure*}

A comparison to nuclear analyses requires the same resonance definition.
For example, the extended $R$-matrix continuation of
Cs\'ot\'o and Hale~\cite{Csoto1997} gives $^5$He poles
$(E_R,\Gamma)=(0.80,0.65)$ MeV and $(2.07,5.57)$ MeV
for $3/2^-$ and $1/2^-$, respectively.
These literature values provide physical context, but are not
raw experimental points with independent covariance.
They also differ from cross-section-peak definitions discussed
in that work. The present model's ODE poles
$(0.76911,0.65114)$ MeV and $(2.00828,5.42406)$ MeV
are of the expected scale, without establishing a new
precision determination of the experimental resonances.
The difference between a QPU pole and its own ODE reference
must not be used as the uncertainty of either literature comparison.

With $x=\cos\theta$ and
$S_{l\pm}=\exp[2i\delta_{l,j=l\pm1/2}]$,
the spin amplitudes in our convention are
\begin{align}
 A(E,x)&=\frac{1}{2ik}\sum_l
 [(l+1)(S_{l+}-1)+l(S_{l-}-1)]P_l(x),\\
 B(E,x)&=\frac{1}{2ik}\sum_{l\geq1}(S_{l+}-S_{l-})P_l^1(x),
\end{align}
where $P_l^1$ follows the Condon--Shortley convention. The angular integral of
\begin{equation}
 \frac{\dd\sigma}{\dd\Omega}=|A|^2+|B|^2
\end{equation}
reproduces the partial-wave sum in the Letter. For the calibrated interaction,
including $l=3,4$ changes the angle-integrated cross section by less than
$0.0066\%$ on the energy grid used in the Letter.

\section{Repeat campaign, shot budget and simulator controls}
Table~\ref{tab:blocks} reports all three QPU blocks.
Each block contains ten physical targets, three arms,
nine queries, two replicates and 512 shots per query and replicate.
This gives 540 circuit evaluations and 276\,480 shots per block.
The three blocks ran on 6 October 2026, with 174 and 63 minutes between
successive execution windows (end to start). The two-qubit and readout
calibrations were unchanged across the three snapshots. The ancilla
$(T_1,T_2)$ estimates changed from $(143.27,199.51)$ to $(280.28,331.12)$
microseconds in the last snapshot. The three blocks share the same device
and ancilla.

\begin{table}[t]
\centering
\caption{Observed follow-up QPU energy RMSEs, in eV.
Each arm has 9216 shots per target and block;
the full sign mean combines two arms and uses 18\,432 shots.
The ten targets in each row share the same execution block.}
\label{tab:blocks}
\begin{tabular}{lrrrr}
\toprule
Block & Forward & Reverse & Half time & Sign mean\\
\midrule
r00 & 52.53 & 57.42 & 135.20 & 19.20 \\
r01 & 54.86 & 47.27 & 132.09 & 13.48 \\
r02 & 40.09 & 60.33 & 132.35 & 17.60 \\
\bottomrule
\end{tabular}
\end{table}

\begin{table*}[t]
\centering
\caption{Channel dependence of follow-up RMSEs in eV.
Each entry includes two trap levels in three blocks.
Forward and reverse each use $N=9216$ shots per target and block;
the full mean uses $2N$. Each equal-shot mean uses one
4608-shot forward replicate and one 4608-shot reverse replicate.}
\label{tab:channels}
\begin{tabular}{lrrrrr}
\toprule
& Forward & Reverse & Full mean & Equal-shot 0 & Equal-shot 1\\
\midrule
$s_{1/2}$ & 96.74 & 69.66 & 17.46 & 23.73 & 15.71 \\
$p_{3/2}$ & 50.68 & 97.12 & 29.53 & 36.19 & 27.68 \\
$p_{1/2}$ & 7.91 & 25.02 & 14.09 & 13.95 & 15.60 \\
$d_{5/2}$ & 15.13 & 10.01 & 4.20 & 8.31 & 5.90 \\
$d_{3/2}$ & 8.75 & 16.55 & 6.39 & 7.03 & 6.99 \\
\bottomrule
\end{tabular}
\end{table*}

The fixed equal-weight mean is evaluated for every target without
subtracting its known residual.
The improvement is substantial for the $s$ wave and both $d$ waves,
but the $p_{1/2}$ forward estimates were already close to their
references and their average with reversed evolution is worse
(Table~\ref{tab:channels}).
The improvement in the pooled RMSE thus depends on the channel.
Both equal-shot partitions are retained to avoid choosing
an allocation after looking at the reference error.
They are not independent repetitions because they come from
the same target designs, device windows and full data set.

A phenomenological decomposition
\begin{equation}
 \widehat E_\pm=E+b_{\rm even}\pm b_{\rm odd}+\eta_\pm
\end{equation}
explains what the sign mean can and cannot remove.
It cancels $b_{\rm odd}$ algebraically but leaves $b_{\rm even}$,
and its variance contains
$[\operatorname{Var}(\eta_+)+\operatorname{Var}(\eta_-)
+2\operatorname{Cov}(\eta_+,\eta_-)]/4$.
An assumption of zero covariance is not justified by merely
submitting the two circuits separately.
The measured sign contrast is consistent with an odd component
of the effective center error, but is not a proof that a specific
coherent gate error causes it.

\begin{table*}[t]
\centering
\caption{Simulator and QPU controls for the follow-up design.
RMSEs in eV include all 30 target--block observations for each arm.
The rejection count refers to all 90 three-arm histogram fits.
The ideal simulator still has finite sampling noise.}
\label{tab:simulators}
\begin{tabular}{lrrrrr}
\toprule
Source & Forward & Reverse & Half time & Sign mean & $p<0.01$\\
\midrule
Ideal & 9.05 & 8.39 & 17.69 & 6.24 & 2/90 \\
Gate noise & 14.38 & 14.82 & 36.56 & 11.52 & 3/90 \\
Backend noise & 11.32 & 13.86 & 27.05 & 8.24 & 2/90 \\
IBM Boston & 49.59 & 55.29 & 133.22 & 16.93 & 72/90 \\
\bottomrule
\end{tabular}
\end{table*}

The synthetic gate-noise configuration uses one- and two-qubit
error probabilities $2\times10^{-4}$ and $0.003$,
reset error $0.001$, and asymmetric readout probabilities
$0.01$ and $0.015$.
The backend-calibrated Aer data use the selected device's
calibration-based noise model; they are local simulations,
not executions on an IBM QPU.
Neither model reproduces the systematic histogram discrepancy
of the hardware (Table~\ref{tab:simulators}).

\begin{figure*}[t]
\centering
\includegraphics[width=\textwidth]{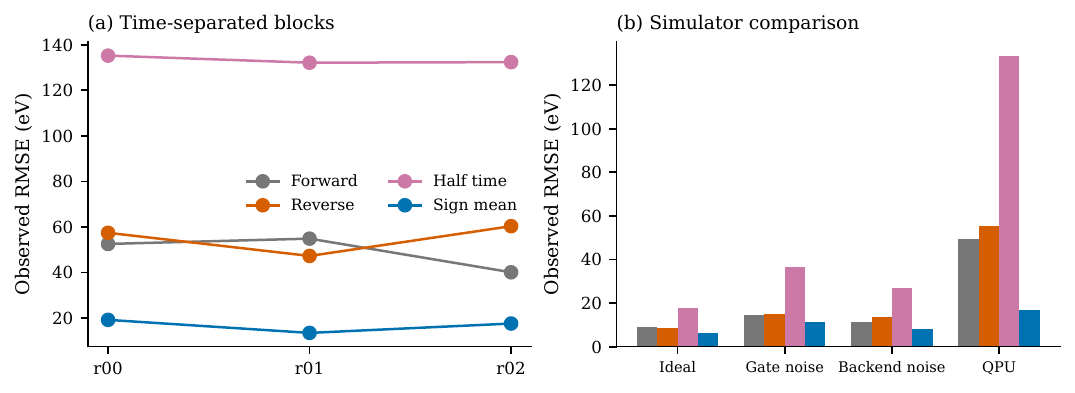}
\caption{Energy RMSEs in the follow-up design.
(a) Block dependence on hardware.
(b) Identical arm comparisons for finite-shot ideal simulation,
synthetic gate noise, calibration-based backend noise and QPU data.
The full sign mean always uses both forward and reverse shot budgets.}
\end{figure*}